\documentclass[12pt]{article}

\usepackage{scicite}

\usepackage{times,ulem}

\newenvironment{sciabstract}{%
\begin{quote} \bf}
{\end{quote}}

\usepackage{graphicx}
\usepackage[colorlinks=true, allcolors=blue]{hyperref}
\usepackage{siunitx}

\title{Spin-wave frequency multiplication by magnetic vortex cores}
\author{
Cheng-Jie Wang,$^{12\dagger}$  Yuxin Li,$^{123\dagger}$ Zhe Ding,$^{123}$\\
Pengfei Wang,$^{123\ast}$ Fazhan Shi,$^{1234\ast}$ Jiangfeng Du$^{1235\ast}$\\
\\
\normalsize{$^{1}$CAS Key Laboratory of Microscale Magnetic Resonance and School of Physical Sciences,}\\
\normalsize{University of Science and Technology of China, Hefei 230026, China}\\
\normalsize{$^{2}$Anhui Province Key Laboratory of Scientific Instrument Development and Application,}\\
\normalsize{University of Science and Technology of China, Hefei 230026, China}\\
\normalsize{$^{3}$Hefei National Laboratory, University of Science and Technology of China, Hefei 230088, China}\\
\normalsize{$^{4}$School of Biomedical Engineering and Suzhou Institute for Advanced Research,}\\
\normalsize{University of Science and Technology of China, Suzhou 215123, China}\\
\normalsize{$^{5}$Institute of Quantum Sensing and School of Physics, Zhejiang University, Hangzhou 310027, China}\\
\\
\normalsize{$^\ast$To whom correspondence should be addressed; }\\
\normalsize{E-mail:  djf@ustc.edu.cn (J.D.); fzshi@ustc.edu.cn (F.S.); wpf@ustc.edu.cn (P.W.).}\\
\normalsize{$^\dagger$These author contributed equally to this work.}
}
\date{}

\begin{document}
\baselineskip24pt
\maketitle

\begin{sciabstract}
Frequency multiplication involves generating harmonics from an input frequency, a technique particularly useful for integrating spin-wave devices operating at different frequencies. While topological magnetic textures offer distinct advantages in spin-wave applications, frequency multiplication has not yet been observed in these structures. Here, we study the magnetization dynamics of magnetic vortices formed in micron-sized disks and squares via wide-field magnetic imaging. We found the occurrence of coherent spin-wave harmonics arising from the gyration of vortex cores driven by microwave fields. This phenomenon reveals a universal mechanism where the periodical motion of delta function-like objects such as vortex cores gives rise to a frequency comb. Our results pave the way for creating nanoscale, tunable spin-based frequency multipliers and open new possibilities for frequency comb generation in a variety of systems.
\end{sciabstract}

\section*{Introduction}

Frequency multiplication is a nonlinear process in which the output signal contains frequency components that are integer multiples of the input frequency. This phenomenon has been extensively studied in various physical systems, including optics\cite{terhune_optical_1962,kleinman_theory_1962,li_ultrahigh-factor_2021} and electronics\cite{raisanen_frequency_1992,siegel_terahertz_2002,chattopadhyay_all-solid-state_2004}. Spintronics is an emerging field that leverages the intrinsic spin of electrons, including phenomena like spin waves, to develop next-generation electronic devices. These spin-based devices offer several advantages such as reduced power consumption and higher operating frequencies\cite{chumak_magnon_2015}. The capability of frequency conversion is particularly desirable to interface spintronics with conventional electronics and communicate between devices operating at different frequencies. Although considerable efforts have been made to demonstrate spin-wave frequency multiplication, they suffer from a limited number of harmonics\cite{hermsdoerfer_spin-wave_2009,demidov_generation_2011,gros_imaging_2022},  the challenge of engineering \cite{koerner_frequency_2022,wu_wideband_2024}, or the difficulty of scaling down to nanoscale dimensions.

To address these issues, one promising approach involves the use of topological magnetic textures, which are characterized by tunability and nanoscale size and provide possibilities for generating spin-wave harmonics\cite{rodrigues_nonlinear_2021}. Among them, magnetic vortices could be ideal candidates for magnonic frequency multipliers because of their abundant spin excitation modes\cite{buess_fourier_2004,perzlmaier_spin-wave_2005,guslienko_vortex-state_2005} and ability to emit short-wavelength spin waves\cite{wintz_magnetic_2016,dieterle_coherent_2019}. However, frequency multiplication via magnetic vortices has not yet been observed in experiments.

In this work, we explore the spin-wave dynamics of magnetic vortices formed in micron-sized disks and squares using a wide-field microscope based on nitrogen-vacancy (NV)  centers\cite{taylor_high-sensitivity_2008,steinert_high_2010}. We observe harmonics extending over 14 orders of the input microwave (MW) frequency and find the correlation between the spatial distributions of these harmonics and vortex structures. Theoretical analysis based on micromagnetic simulations reveals that the frequency multiplication process originates in the gyration of magnetic vortex cores. In addition, the harmonics are proven to be phase-stable by driving Rabi oscillations of NV centers.

\section*{Results}

In our experiments, micron-sized $\rm{Ni_{80}Fe_{20}}$ disks and squares are fabricated on the surface of a diamond chip containing a shallow layer of dense NV centers, and covered with Au striplines as microwave antennas, as illustrated in Fig.\ref{fig:introduction}A (further details in Supplementary Materials). The MW field generated by the stripline excites the magnetization dynamics in disks and squares as well as the electron spin resonance (ESR) in NV centers. The ESR causes a reduction in the photoluminescence (PL) intensity and thus is optically detected, i.e., optically detected magnetic resonance (ODMR). 
A magnetic field $B$ splits the ground states of NV spins into two resonance frequencies $f_\mathrm{NV}=D\pm\gamma B$, where $D=2.87~\rm{GHz}$ is the zero-field splitting and $\gamma=28~\mathrm{MHz/mT}$ is the gyromagnetic ratio (see Fig.\ref{fig:introduction}B). In this way, stray fields of magnetic structures can be imaged by mapping $f_\mathrm{NV}$. Moreover, the dynamic stray fields generated by coherent spin waves can excite ODMR signals in the same way as MW fields, which enables NV centers to image spin waves\cite{van_der_sar_nanometre-scale_2015,bertelli_magnetic_2020,lee-wong_nanoscale_2020,mccullian_broadband_2020}.

In Fig.\ref{fig:introduction}C, the stray field imaged with NV centers shows a typical vortex configuration, where only the vortex core yields a relatively large value at the center of the disk\cite{tetienne_quantitative_2013}. The measured PL change of NV centers is shown as a function of the MW frequency $f_{\rm{MW}}$ in Fig.\ref{fig:introduction}D. The resonance frequency of NV centers is 2.87 GHz and the resonance frequency of the excited states is also detected at approximately 1.42 GHz. In addition to these ESR signals of NV centers (dashed lines), a strong response at approximately 1.9 GHz (the dotted line) and sharp peaks are observed close to the micro-disk. These additional signals are similar to those observed in magnetic thin films\cite{koerner_frequency_2022}. The enhancement at $2/3~f_\mathrm{NV}\approx 1.9~\mathrm{GHz}$ can be attributed to nonlinear spin-wave excitation\cite{bauer_nonlinear_2015}. Importantly, sharp peaks are identified at $f_{\rm{MW}}=f_\mathrm{NV}/n$, where $n$ is an integer, and are the results of a frequency multiplication process. This process has been attributed to magnetization switching between two energetically equivalent states at the boundary of magnetic textures in magnetic thin films. For a micro-disk, the demagnetized field aligns the magnetization circumferentially except near the disk center. This effect is known as geometric anisotropy and is the cause of vortex structures. In this scenario, it is unlikely for energetically equivalent states to exist, similar to the case where the magnetization is fully saturated by in-plane bias fields. Therefore, the observed phenomenon is likely driven by an undiscovered mechanism.

\begin{figure}
    \centering
    \includegraphics[width=1\linewidth]{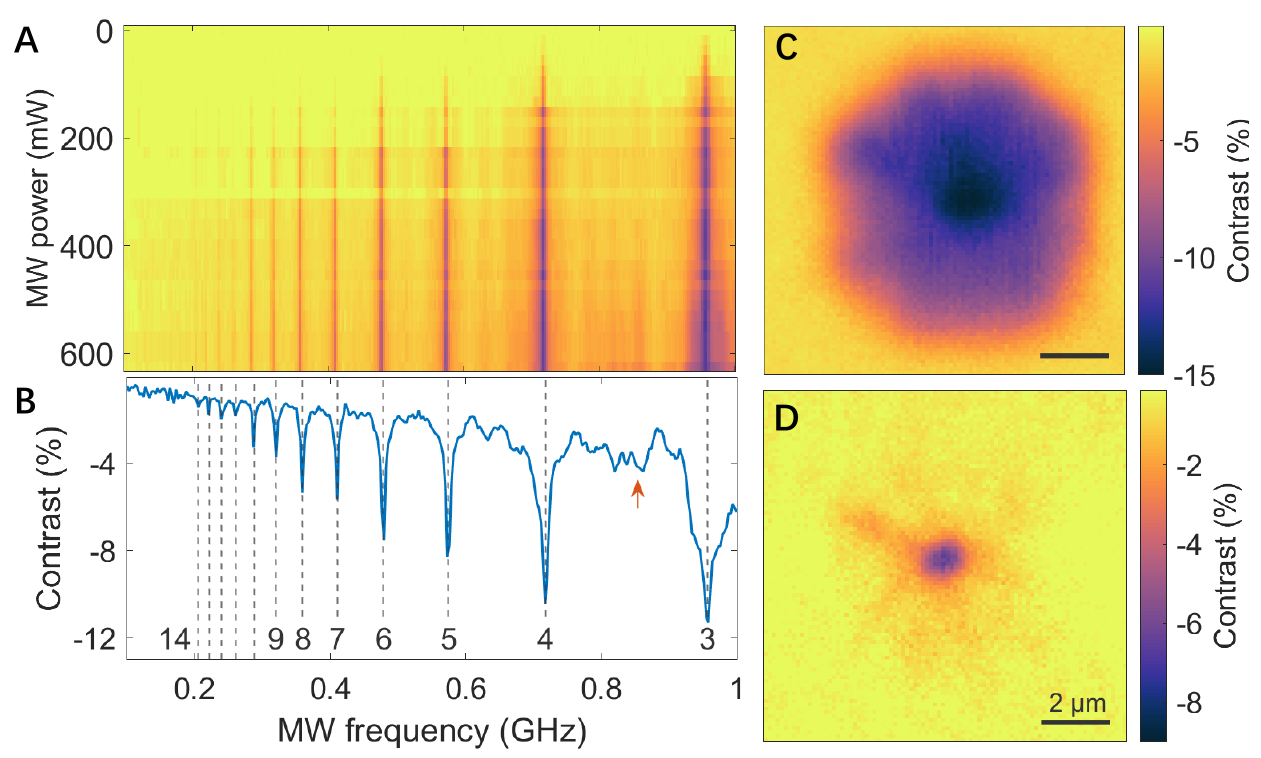}
    \caption{Harmonics of the excitation frequency. (A) PL contrast as a function of MW power and frequency. The MW power spans a range from -12 dBm to 28dBm. Note that he MW power mentioned in this article refers to the power input to the antenna connector rather than the stripline. (B) Scan of the MW frequency under an MW power of 28 dBm. The dashed lines denote the positions of $f_\mathrm{NV}/n,~n=\numrange[range-phrase = -]{3}{14}$. (C) Intensity map of the 4th harmonic. The contrast is the maximum of the 4th harmonic peak. (D) Spatial distribution of the PL contrast at 852.5 MHz, denoted by the red arrow in (B). (C) and (D) are measured at 28 dBm, corresponding to an MW field of 409 \si{\micro\tesla}. The data in (A) and (B) are averaged over the entire disk region.}
    \label{fig:harmonics}
\end{figure}

Next, we measure the ODMR spectrum under different excitation powers. In Fig.\ref{fig:harmonics}A, it can be seen that harmonics intensify and higher harmonics emerge as the excitation power increases. Under a high excitation power,  harmonics up to the 14th order can be clearly observed as shown in Fig.\ref{fig:harmonics}B. Besides, in both Fig.\ref{fig:harmonics}A and B, higher-order harmonics remain detectable despite the relatively low signal-to-noise ratio.
To investigate the origin of these harmonics, the intensity map of the 4th harmonic is displayed in Fig.\ref{fig:harmonics}C. We note that the harmonic exhibits the strongest response at the center of the disk. Also, we observe a weak response at the disk center when multiples of the excitation frequency are not resonant with NV centers ,i.e., $f_{\mathrm{MW}}\neq f_\mathrm{NV} /n$  (Fig.\ref{fig:harmonics}D). This phenomenon is not comprehensible within the context of the switching model. These findings suggest that the generation of harmonics is related to magnetic vortex structures. This assumption is further supported by the disappearance of harmonics when we apply in-plane bias fields to disrupt vortex structures (fig.S2).

In addition to measurements conducted with disks, we perform same measurements with magnetic vortices formed in squares (more results are provided in Supplementary Materials). Fig.\ref{fig:sqaure}A shows the stray field of a typical vortex structure in squares\cite{rondin_stray-field_2013}. While the ODMR spectrum measured with squares (Fig.\ref{fig:sqaure}B) is similar to the frequency comb observed with disks, intensity maps of harmonics reflect the domain structure in squares. As can be seen in Fig.\ref{fig:sqaure}C, high intensity is distributed along the domain walls in spite of the vortex core. This may provide a method to control the propagation of spin-wave harmonics because magnetic domain walls have been exploited as spin-wave nanochannels\cite{wagner_magnetic_2016}.

\begin{figure}
    \centering
    \includegraphics[width=1\linewidth]{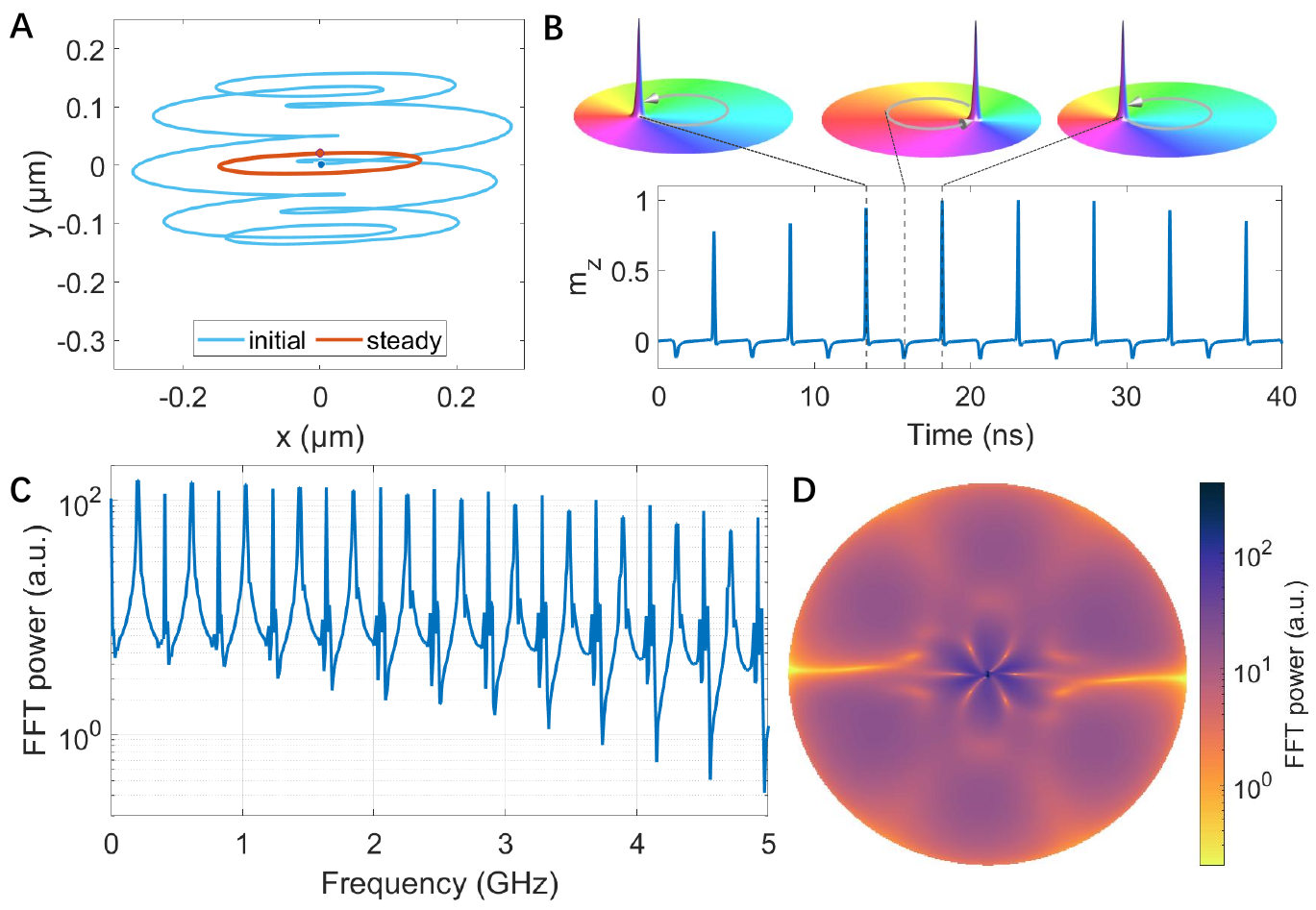}
    \caption{Micromagnetic simulations and vortex core gyration. (A) Vortex core trajectories driven by MW fields. The blue dot denotes the equilibrium position of the vortex core. The blue line and red line are trajectories in the initial state and steady state, respectively. (B) Evolution of $m_\mathrm{z}$. The upper part demonstrates the distributions of $m_\mathrm{z}$ at different moments. (C) Fourier spectrum of $m_\mathrm{z}$. (D) Intensity map of the 3rd harmonic excited at 900 MHz.}
    \label{fig:simulation}
\end{figure}

To reveal the physical mechanism of the frequency multiplication process, we perform micromagnetic simulations to reproduce the spin dynamics in magnetic vortices (details in Supplementary Materials). It is well known that the vortex core experiences a gyration when driven by an MW field and its period is determined by the MW frequency. If the MW frequency is resonant with the gyrotropic eigen-mode, which is 30 MHz for the disks in our simulations, the gyration orbit is circular in the steady state or spiral in the transient state. However, the vortex core is driven by MW fields with arbitrary frequencies in our experiments. In this case, vortex-core trajectories are elliptical or stadium-like in the steady state\cite{lee_gyrotropic_2007}.  
The vortex-core trajectories driven by an MW field with a frequency of 205 MHz, which is $f_\mathrm{NV}/14$, are shown in Fig.\ref{fig:simulation}A. As can be seen, the trajectory is initially complex (blue line) and then gradually approaches an ellipse (red line). It has been reported that the gyration of vortex cores can generate spin waves\cite{wintz_magnetic_2016,dieterle_coherent_2019}. To analyse the spectrum of spin waves generated by the vortex core, the evolution of the out-plane normalized magnetization $m_\mathrm{z}$ in the elliptical orbit (red dot in Fig.\ref{fig:simulation}A) is extracted and displayed in Fig.\ref{fig:simulation}B. The evolution of $m_\mathrm{z}$ is characterized by periodic spikes because $m_\mathrm{z}$ at the orbit undergoes a periodic rapid change carried by the vortex core. In other words, the $m_\mathrm{z}$ distribution of the vortex core approximately follows a Dirac delta function, which is illustrated in the upper part of Fig.\ref{fig:simulation}B. As a result, the gyration of the vortex core causes the evolution of $m_\mathrm{z}$ to form a Dirac comb. This behavior clearly explains the appearance of frequency multiplication. As expected, the spectrum of $m_\mathrm{z}$ is a frequency comb (Fig.\ref{fig:simulation}C). To summarize, this frequency multiplication process benefits from two characteristics of magnetic vortices: a delta function-like vortex core and its gyration period determined by the excitation frequency.

Meanwhile, the movement of the vortex core induces propagating perturbations, serving as a source of spin waves. Hence, the magnetization evolution away from the orbit can also form a Dirac comb with low intensity (fig.S3). The intensity map of the 3rd harmonic is displayed in Fig.\ref{fig:simulation}D and other orders are provided in Supplementary Materials. As can be seen, the high intensity at the center is in good agreement with the experimental results and can be easily understood as a consequence of the vortex core orbit. As for the flower-shaped distribution, it may be assumed that the distribution is distorted by polycrystalline grain structure, which is demonstrated via simulations (fig.S4), or the inhomogeneity of the magnetic thin film caused by imperfect fabrication.

It is noteworthy that the vortex core's capacity to generate harmonics, despite moving as a linear oscillator, stems from the inherent non-linear characteristics of its magnetic texture. In essence, while the vortex core trajectory exhibits harmonic oscillation, its magnetization undergoes an an-harmonic evolution. Particularly, a delta function-like field in harmonic oscillation manifests as a pulse train in the temporal domain at each point along the trajectory and generates a frequency comb (see Supplementary Materials for general case). We note that the nonlinear behavior of vortex oscillations has been demonstrated in nanocontacts \cite{pufall_low-field_2007,mistral_current-driven_2008,petit-watelot_commensurability_2012,devolder_chaos_2019}, although it's not a frequency multiplication process and not observation of spin waves. 

In addition, above analysis is not relevant to vortex core reversal dynamics. In fact, the excitation field in this work is insufficient to induce core reversal because the excitation frequency is far from the gyrotropic eigen-frequency\cite{lee_universal_2008}. Specifically, core reversal is not observed under an excitation field of 1 mT in the simulation shown in Fig.\ref{fig:simulation} . It can be inferred that vortex core reversal could not happen in our experiments since MW fields are less than 1 mT.

\begin{figure}
    \centering
    \includegraphics[width=0.5\linewidth]{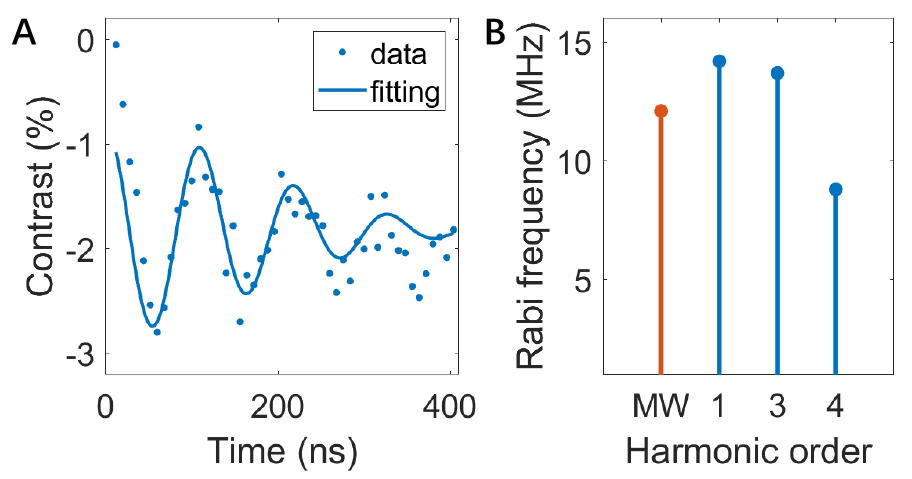}
    \caption{Rabi oscillations driven by harmonics. (A) Rabi signal driven by the 3rd harmonic, i.e. \(f_\mathrm{MW}=0.9~\mathrm{GHz}\). The line is fitted via a cosine function coupled with an exponential decay.  The signal was measured under an MW power of 28 dBm, corresponding to 409 \si{\micro\tesla}.(B) Rabi frequency variation with different harmonics under the same excitation power. The Rabi frequency measured far from the disk is displayed as a reference (red line). These Rabi oscillations are displayed in fig.S8.}
    \label{fig:rabi}
\end{figure}

To evaluate the potential application of the frequency multiplication process caused by magnetic vortices, we use harmonics to drive the Rabi oscillation of NV centers. Although Rabi oscillations measured with NV ensembles are disrupted by the gradient of stray fields (fig.S7), distinct Rabi signals can be observed as shown in Fig.\ref{fig:rabi}A. This capability of executing quantum coherent control substantiates that the harmonics produced by the vortex core gyration are phase-stable.
Because the Rabi frequency can precisely represent the amplitude of dynamic magnetic fields, it provides a quantitative tool for investigating the harmonic excitation effect. In Fig.\ref{fig:rabi}B, we compare the Rabi oscillations driven by different harmonics under identical excitation powers. Although the frequency conversion efficiency gradually decreases with increasing harmonic order, the 3rd-order harmonic retains a high efficiency compared with stripline-generated MW fields. This result implies an enhancement induced by magnetic thin films as reported previously\cite{wolf_strong_2017,wang_electrical_2020}.

\section*{Discussion and Conclusion}

Our experiments present the occurrence of frequency multiplication extending over 14 orders in magnetic vortices. Based on intensity maps of harmonics and micromagnetic simulations, we attribute the origin of the frequency multiplication process to spin-wave dynamics induced by the gyration of vortex cores. The harmonics generated by this process are proven to be phase-stable because of their capability to drive Rabi oscillations of NV centers. This feature provides support for the potential applications of the frequency multiplication technique. In particular, the possibility of qubit manipulation using an off-resonant input frequency opens new perspectives for quantum information\cite{wang_electrical_2020,bejarano_parametric_2024,wu_wideband_2024}. Although the excitation efficiency is in-homogeneous in our experiments, improvement can be achieved by refining fabrication techniques or employing alternative magnetic materials such as yttrium iron garnet (YIG). 

One significant advantage of magnetic vortices is their ability to excite short-wavelength spin waves by magnetic vortex cores\cite{wintz_magnetic_2016,dieterle_coherent_2019}. The wavelength of the harmonic spin waves observed in our experiments should be measured in future work but it can be estimated with our results. In fact, NV sensors can act as a wavelength filter defined by the probe-sample distance, because dynamic stray fields generated by spin waves decay evanescently on the scale of the spin-wavelength\cite{van_der_sar_nanometre-scale_2015,casola_probing_2018,simon_filtering_2022}. In our experiments, the probe-sample distance of approximately 100 nm gives rise to a wavelength filter peaking at 100 nm. Therefore, harmonics measured by NV centers can be assumed to have a wavelength below a micron on account of their high-contrast ODMR signals.
Furthermore, our experiments demonstrate the well-established advantages of magnetic vortices, including their size, tunability and stability: frequency multiplication is realized in micron-sized disks (squares) and simulated in a disk with a diameter of 500 nm (fig.S10); tuning harmonic behaviors by the geometric shape is demonstrated with squares; all these advantages are achieved without the need for a bias field. These results underscore the potential of magnetic vortices in the development of practical spin-based frequency multipliers.

From a basic perspective, rapidly changing objects occurring within a limited space can potentially behave like a Dirac delta function. When these objects are driven to move periodically, they can generate a pulse train in the time domain and thus a comb structure in the frequency domain. In addition to vortex cores, this type of objects can include solitons or singularities such as domain walls (one-dimensional delta function along the normal direction) and Bloch points in magnetic systems. For example, this framework offers an intuitive explanation for the multiple harmonics generated by domain wall oscillations\cite{zhou_spin_2021,wu_wideband_2024}. Furthermore, the vortex structure and other delta function-like objects are ubiquitous across various types of systems such as optics or superconductors. For example, dissipative solitons have been used to generate optical frequency combs \cite{herr_temporal_2014,kippenberg_dissipative_2018}. Therefore, our model could provide valuable insights for frequency comb generation in various systems.

\bibliographystyle{Science}

\begin{thebibliography}{10}

    \bibitem{terhune_optical_1962}
    R.~W. Terhune, P.~D. Maker, C.~M. Savage, {\it Phys. Rev. Lett.\/} {\bf 8}, 404
      (1962). Publisher: American Physical Society.
    
    \bibitem{kleinman_theory_1962}
    D.~A. Kleinman, {\it Phys. Rev.\/} {\bf 128}, 1761 (1962). Publisher: American
      Physical Society.
    
    \bibitem{li_ultrahigh-factor_2021}
    J.~Li, {\it et~al.\/}, {\it Opt. Express, OE\/} {\bf 29}, 40748 (2021).
      Publisher: Optica Publishing Group.
    
    \bibitem{raisanen_frequency_1992}
    A.~Raisanen, {\it Proceedings of the IEEE\/} {\bf 80}, 1842 (1992). Conference
      Name: Proceedings of the IEEE.
    
    \bibitem{siegel_terahertz_2002}
    P.~Siegel, {\it IEEE Transactions on Microwave Theory and Techniques\/} {\bf
      50}, 910 (2002). Conference Name: IEEE Transactions on Microwave Theory and
      Techniques.
    
    \bibitem{chattopadhyay_all-solid-state_2004}
    G.~Chattopadhyay, {\it et~al.\/}, {\it IEEE Transactions on Microwave Theory
      and Techniques\/} {\bf 52}, 1538 (2004). Conference Name: IEEE Transactions
      on Microwave Theory and Techniques.
    
    \bibitem{chumak_magnon_2015}
    A.~V. Chumak, V.~I. Vasyuchka, A.~A. Serga, B.~Hillebrands, {\it Nat. Phys.\/}
      {\bf 11}, 453 (2015).
    
    \bibitem{hermsdoerfer_spin-wave_2009}
    S.~J. Hermsdoerfer, {\it et~al.\/}, {\it Applied Physics Letters\/} {\bf 94},
      223510 (2009).
    
    \bibitem{demidov_generation_2011}
    V.~E. Demidov, {\it et~al.\/}, {\it Phys. Rev. B\/} {\bf 83}, 054408 (2011).
      Publisher: American Physical Society.
    
    \bibitem{gros_imaging_2022}
    F.~Groß, {\it et~al.\/}, {\it Phys. Rev. B\/} {\bf 106}, 014426 (2022).
      Publisher: American Physical Society.
    
    \bibitem{koerner_frequency_2022}
    C.~Koerner, {\it et~al.\/}, {\it Science\/} {\bf 375}, 1165 (2022). Publisher:
      American Association for the Advancement of Science.
    
    \bibitem{wu_wideband_2024}
    J.~Wu, {\it et~al.\/}, {\it npj Spintronics\/} {\bf 2}, 1 (2024). Publisher:
      Nature Publishing Group.
    
    \bibitem{rodrigues_nonlinear_2021}
    D.~Rodrigues, {\it et~al.\/}, {\it Phys. Rev. Appl.\/} {\bf 16}, 014020 (2021).
      Publisher: American Physical Society.
    
    \bibitem{buess_fourier_2004}
    M.~Buess, {\it et~al.\/}, {\it Phys. Rev. Lett.\/} {\bf 93}, 077207 (2004).
    
    \bibitem{perzlmaier_spin-wave_2005}
    K.~Perzlmaier, {\it et~al.\/}, {\it Phys. Rev. Lett.\/} {\bf 94}, 057202
      (2005). Publisher: American Physical Society.
    
    \bibitem{guslienko_vortex-state_2005}
    K.~Y. Guslienko, W.~Scholz, R.~W. Chantrell, V.~Novosad, {\it Phys. Rev. B\/}
      {\bf 71}, 144407 (2005). Publisher: American Physical Society.
    
    \bibitem{wintz_magnetic_2016}
    S.~Wintz, {\it et~al.\/}, {\it Nature Nanotechnology\/} {\bf 11}, 948 (2016).
    
    \bibitem{dieterle_coherent_2019}
    G.~Dieterle, {\it et~al.\/}, {\it Phys. Rev. Lett.\/} {\bf 122}, 117202 (2019).
    
    \bibitem{taylor_high-sensitivity_2008}
    J.~M. Taylor, {\it et~al.\/}, {\it Nat. Phys.\/} {\bf 4}, 810 (2008).
    
    \bibitem{steinert_high_2010}
    S.~Steinert, {\it et~al.\/}, {\it Review of Scientific Instruments\/} {\bf 81},
      043705 (2010).
    
    \bibitem{van_der_sar_nanometre-scale_2015}
    T.~van~der Sar, F.~Casola, R.~Walsworth, A.~Yacoby, {\it Nat. Commun.\/} {\bf
      6}, 7886 (2015).
    
    \bibitem{bertelli_magnetic_2020}
    I.~Bertelli, {\it et~al.\/}, {\it Science Advances\/} {\bf 6}, eabd3556 (2020).
      Publisher: American Association for the Advancement of Science Section:
      Research Article.
    
    \bibitem{lee-wong_nanoscale_2020}
    E.~Lee-Wong, {\it et~al.\/}, {\it Nano Lett.\/} {\bf 20}, 3284 (2020).
      Publisher: American Chemical Society.
    
    \bibitem{mccullian_broadband_2020}
    B.~A. McCullian, {\it et~al.\/}, {\it Nature Communications\/} {\bf 11}, 5229
      (2020). Number: 1 Publisher: Nature Publishing Group.
    
    \bibitem{tetienne_quantitative_2013}
    J.-P. Tetienne, {\it et~al.\/}, {\it Phys. Rev. B\/} {\bf 88}, 214408 (2013).
    
    \bibitem{bauer_nonlinear_2015}
    H.~G. Bauer, P.~Majchrak, T.~Kachel, C.~H. Back, G.~Woltersdorf, {\it Nat
      Commun\/} {\bf 6}, 8274 (2015). Publisher: Nature Publishing Group.
    
    \bibitem{rondin_stray-field_2013}
    L.~Rondin, {\it et~al.\/}, {\it Nat. Commun.\/} {\bf 4}, 2279 (2013).
    
    \bibitem{wagner_magnetic_2016}
    K.~Wagner, {\it et~al.\/}, {\it Nature Nanotech\/} {\bf 11}, 432 (2016).
    
    \bibitem{lee_gyrotropic_2007}
    K.-S. Lee, S.-K. Kim, {\it Applied Physics Letters\/} {\bf 91}, 132511 (2007).
    
    \bibitem{pufall_low-field_2007}
    M.~R. Pufall, W.~H. Rippard, M.~L. Schneider, S.~E. Russek, {\it Phys. Rev.
      B\/} {\bf 75}, 140404 (2007). Publisher: American Physical Society.
    
    \bibitem{mistral_current-driven_2008}
    Q.~Mistral, {\it et~al.\/}, {\it Phys. Rev. Lett.\/} {\bf 100}, 257201 (2008).
      Publisher: American Physical Society.
    
    \bibitem{petit-watelot_commensurability_2012}
    S.~Petit-Watelot, {\it et~al.\/}, {\it Nature Phys\/} {\bf 8}, 682 (2012).
      Publisher: Nature Publishing Group.
    
    \bibitem{devolder_chaos_2019}
    T.~Devolder, {\it et~al.\/}, {\it Phys. Rev. Lett.\/} {\bf 123}, 147701 (2019).
      Publisher: American Physical Society.
    
    \bibitem{lee_universal_2008}
    K.-S. Lee, {\it et~al.\/}, {\it Phys. Rev. Lett.\/} {\bf 101}, 267206 (2008).
      Publisher: American Physical Society.
    
    \bibitem{wolf_strong_2017}
    M.~S. Wolf, R.~Badea, M.~Tader, J.~Berezovsky, {\it Physical Review B\/} {\bf
      96} (2017).
    
    \bibitem{wang_electrical_2020}
    X.~Wang, {\it et~al.\/}, {\it npj Quantum Inf\/} {\bf 6}, 1 (2020). Number: 1
      Publisher: Nature Publishing Group.
    
    \bibitem{bejarano_parametric_2024}
    M.~Bejarano, {\it et~al.\/}, {\it Science Advances\/} {\bf 10}, eadi2042
      (2024). Publisher: American Association for the Advancement of Science.
    
    \bibitem{casola_probing_2018}
    F.~Casola, T.~v.~d. Sar, A.~Yacoby, {\it Nat. Rev. Mater.\/} {\bf 3}, 17088
      (2018).
    
    \bibitem{simon_filtering_2022}
    B.~G. Simon, {\it et~al.\/}, {\it Nano Lett.\/} {\bf 22}, 9198 (2022).
      Publisher: American Chemical Society.
    
    \bibitem{zhou_spin_2021}
    Z.-w. Zhou, X.-g. Wang, Y.-z. Nie, Q.-l. Xia, G.-h. Guo, {\it Journal of
      Magnetism and Magnetic Materials\/} {\bf 534}, 168046 (2021).
    
    \bibitem{herr_temporal_2014}
    T.~Herr, {\it et~al.\/}, {\it Nature Photon\/} {\bf 8}, 145 (2014). Publisher:
      Nature Publishing Group.
    
    \bibitem{kippenberg_dissipative_2018}
    T.~J. Kippenberg, A.~L. Gaeta, M.~Lipson, M.~L. Gorodetsky, {\it Science\/}
      {\bf 361}, eaan8083 (2018). Publisher: American Association for the
      Advancement of Science.
    
    \bibitem{vansteenkiste_design_2014}
    A.~Vansteenkiste, {\it et~al.\/}, {\it AIP Advances\/} {\bf 4}, 107133 (2014).
    
    \end{thebibliography}

\section*{Acknowledgments}

This work was supported by the National Natural Science Foundation of China (grant No. T2125011, 11874338), the CAS (grant No. ZDZBGCH2021002, GJJSTD20200001, YSBR-068), Innovation Program for Quantum Science and Technology (Grant No. 2021ZD0302200, 2021ZD0303204), New Cornerstone Science Foundation through the XPLORER PRIZE, the Fundamental Research Funds for the Central Universities, and the USTC Tang Scholar.

This work was partially carried out at the USTC Center for Micro and Nanoscale Research and Fabrication. 

\section*{Supplementary materials}
Methods\\
Figs. S1 to S10\\
References \textit{\cite{vansteenkiste_design_2014}}

\clearpage

\begin{figure}
\centering
\includegraphics[width=0.75\linewidth]{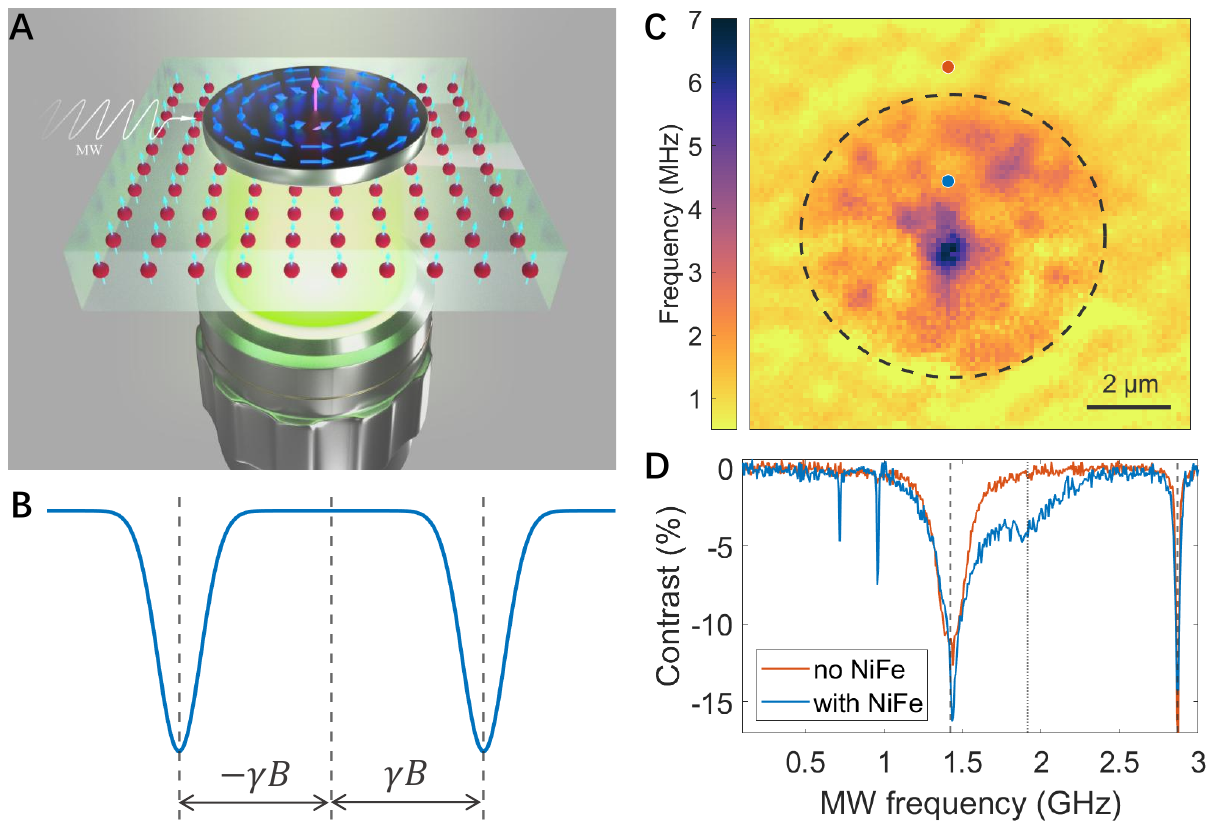}
\caption{\label{fig:introduction} Measurement via wide-field NV microscopy. (A) Experimental setup. The wide-field NV microscope consists of a wide-field optical microscope and a diamond chip containing spin sensors. The $\rm{Ni_{80}Fe_{20}}$ thin film has a thickness of 20 nm. Disks have a diameter of 6.8 \si{\um} and squares have a side-length of 7 \si{\um}. The arrows depict the magnetization configuration of a magnetic vortex. The stripline delivers an in-plane MW field. (B) Schematic spectrum of NV centers. (C) The frequency shift caused by the stray field of the micro-disk. Note that the small stray field produced by the vortex cannot split the ESR peak measured with NV ensembles. The dashed circle denotes the shape of the disk. (D) Dependence of the PL contrast on the MW frequency. The red (blue) line is measured at the position marked by the red (blue) dot in (C). The dashed lines denote the ESR signals of NV centers. The dotted line denotes the enhancement at 1.9 GHz. We note that this enhancement shifts to lower frequencies in some areas (fig.S1). (C) and (D) are measured under a zero bias field.}
\end{figure}

\begin{figure}
    \centering
    \includegraphics[width=0.75\linewidth]{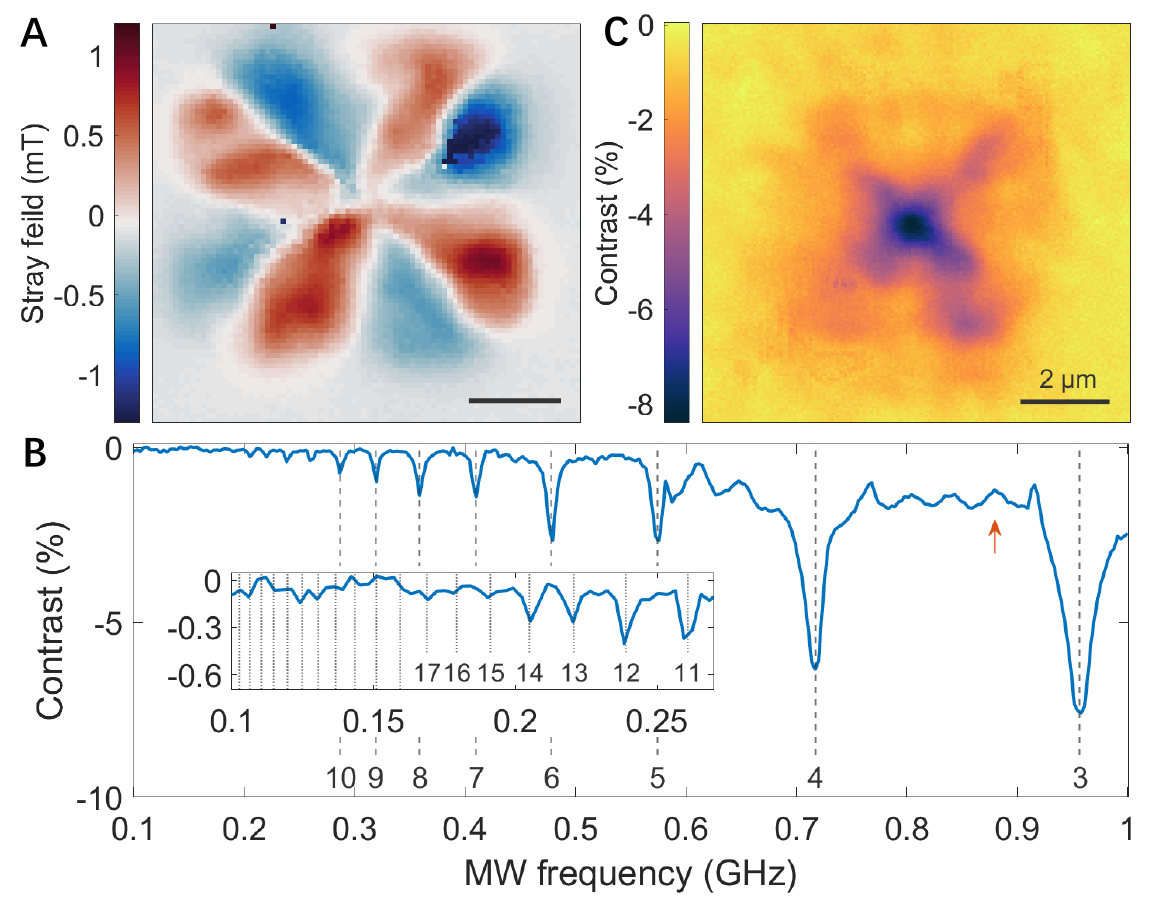}
    \caption{Measurements with squares. (A) Stray field of a square. (B) ODMR spectrum. The data are averaged over the entire square region. The spectrum was measured under an MW power of 21 dBm, corresponding to 183 \si{\micro\tesla}.  (C) Spatial distribution of the PL contrast at 880 MHz, denoted by the red arrow in (B). (A) and (C) are measured under an out-plane bias field of 6.1 mT.}
    \label{fig:sqaure}
\end{figure}

\end{document}